# Micro/nanoscale spacers for enhanced thermophotovoltaic and thermionic energy conversion: a comprehensive review

Nicolas A. Loubet [(a)*], Katie Bezdjian [(a)], Esther López [(a)], Alejandro Datas[(a)]

(a) Instituto de Energía Solar – Universidad Politécnica de Madrid, Avenida Complutense 30, 28040, Madrid, Spain

* Corresponding author: nicolas.loubet@upm.es

## ABSTRACT

Thermionics and thermophotovoltaics are solid-state technologies that convert high-temperature heat into electricity by utilizing fundamental particles—electrons in thermionics and photons in thermophotovoltaics—as energy carriers. Both systems have the potential to achieve high efficiency and power density, contingent on the optimization of radiative/electronic energy fluxes. A critical factor in enhancing energy flux in these devices is the introduction of microscale (thermionics) or nanoscale (thermophotovoltaics) gaps between the hot thermal emitter and the cooler receiver. In thermionic converters, microscale gaps mitigate space charge effects that create energy barriers to electron flow. For thermophotovoltaic systems, nanoscale gaps facilitate photon tunneling, significantly boosting photon flux towards the thermophotovoltaic cell. Forming these small-scale gaps often necessitates intermediate materials or spacers between the emitter and receiver. Over the past few decades, various spacer designs have been proposed and studied, demonstrating their effectiveness in enhancing energy transfer and conversion. However, challenges remain regarding their reliability and scalability. This article provides a comprehensive overview of spacer technologies for thermionics and thermophotovoltaics and summarizes recent advancements, current capabilities, and persistent challenges.

## 1. INTRODUCTION

Thermal to electric energy conversion is the source of most of the world's power. Heat used for energy conversion can be extracted from fossil fuels [1] and nuclear reactions [2], in addition to renewable sources such as concentrated solar power systems [3]. Another potential source is heat produced by industrial processes, as two thirds of the primary energy consumption from fossil fuels is lost, mostly as waste heat [4], [5], [6].

Additionally, heat storage systems coupled with thermophotovoltaic devices present an alternative to electrochemical batteries [7], [8]. Dynamic systems such as steam and gas turbines are currently the most common thermal-to-electric energy converters, but they are cost effective only on a large scale. Solid state devices present a promising modular alternative as their efficiencies do not depend on size, subsequently allowing for the recovery of heat scattered in homes, vehicles, industry, etc. Solid state converters are of special interest as a solution to energy storage because they can be implemented in relatively small systems when compared to dynamic systems. They also offer various advantages compared to traditional converters, such as noiseless operation, easy maintenance, high power density, and scalability [9], [10], [11].

Solid state converters are comprised of thermoelectric generators (TEG), thermionic converters (TIC) and thermophotovoltaic converters (TPV). In TEGs, a temperature gradient drives a flow of electrons in a solid [12], the electron flux is constrained by heat conduction occurring through the medium, which limits the temperature gradient achievable. As a result, the efficiency of TEGs devices is limited to values below 15% [13]. By contrast, TPVs [14], [15] and TICs [16] produce electricity by relying on photons and electrons, respectively, flowing from an emitter to a receiver that are separated by a space. The presence of a gap between the thermal emitter and the receiver enables larger temperature gradients between the hot and cold sides, resulting in higher conversion efficiencies. Currently, the highest experimental conversion efficiency of TPVs is in the range of 25% to 44% [17], [18], [19], [20], [21], while it is around 15% for TICs [22], [23]. Despite the higher efficiency of TIC and TPV devices, TEGs are the most deployed type of converters [24] because they can operate at temperatures below 800 ºC, contrary to TPVs and TICs which need elevated temperatures (>1000 ºC) to reach high power densities and conversion efficiencies. So far, to achieve power densities over 1 $W.cm^{-2}$, emitter temperatures above 1300 K are needed in TICs [22], and above 1500 K in TPVs [25], [26]. However, it is expected that both of these technologies reach higher efficiencies and power densities by optimizing their design and materials [22], [27].

In addition to a heat source, both TPV and TIC devices feature the same basic elements, namely a hot emitter and a cold receiver separated by a vacuum gap. A basic TPV device, depicted in Figure 1, consists of two main parts: 1) a thermal emitter that is typically at a temperature of 1000–2000 K; and 2) a photovoltaic (PV) cell [14], [15]. The external heat source heats the emitter, which then emits thermal radiation towards the photovoltaic cell with an appropriate bandgap energy, converting the radiation into electricity. More advanced TPV devices can also include elements that can achieve spectral control to optimize radiation exchange such as a selective emitter, filters in the gap and mirrors on the cell. A conventional TIC, illustrated in Figure 1, consists of similar parts: a cathode that emits electrons once heated to temperatures above 1000°C [28], [29] and a cold anode, usually comprised of a low work function material, that collects the electrons. Fundamentally, the kinetic energy of the cathode's electrons increases as the cathode is heated by an external source until they are able to overcome the cathode's work function and escape from the surface [28] [30]. If the cathode and anode are connected by an electrical load, the voltage difference drives a current, and electric power is produced [11].

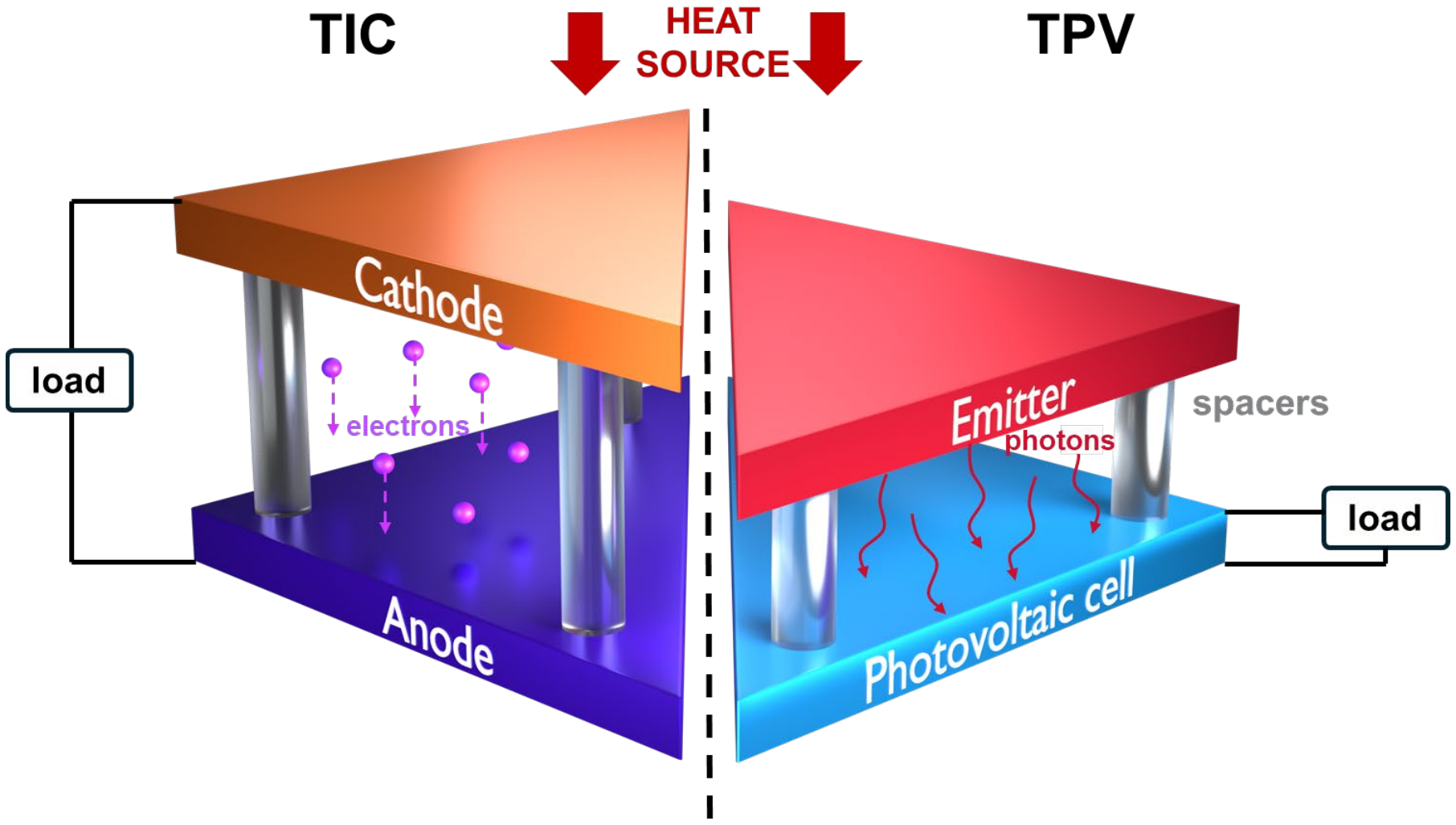

*Figure 1 – Basic components of thermionic and thermophotovoltaic energy converters, inspired by [22]. The gap between the emitter and receiver is enabled by spacers, represented by the cylindrical posts.*

## 2. STATE OF THE ART OF THERMIONIC AND THERMOPHOTOVOLTAIC DEVICES

### 2.1. THERMIONIC CONVERTERS

Thermionic emission was first noted by Thomas Edison in 1885 when he observed that electrons flowed between two electrodes at different temperatures when separated by a vacuum [31]. Although the first thermionic device was proposed by W. Schlichter in 1915 [32], the development of TICs was minimal until the 1950s when the space race necessitated efficient and compact electricity sources for aerospace missions [5], [32]. The first TIC device was achieved by Hatsopoulos and Kaye in 1958 [16]. Due to the limitations of fabrication techniques, research was minimal from the 90s onwards until progress in concentrated solar power and the ability to use sunlight to exploit thermal energy reignited interest in TICs [32]. In practice, however, two main issues hinder the performance of TIC devices: 1) the development of stable materials with a low work function, which are required to fabricate the cathode and anode [33]; and 2) the negative space charge created by electrons that accumulate between the cathode and anode [11], [22], [32], [33]. This cloud of electrons repels electrons emitted by the cathode away from the anode, preventing electrons from reaching the anode and, consequently, reducing electric power production [11]. Researchers have attempted to mitigate the space charge effect through several methods, including altering the material properties of the cathode surface [11], introducing a plasma in the gap [34] or simply reducing the distance [16], [35]. Historically, cesium plasma has been used to suppress the space charge effect, but the addition of plasma highly complicates TICs, and the ionization of cesium consumes energy during the conversion process, diminishing the overall system efficiency by as much as 30–50% [11], [22]. Another approach to reduce the space charge effect is adding a gate or grid between the two electrodes to tune the electric field and accelerate the flow of electrons [36], [37]. Gates or grids and cesium plasma are advantageous because neither method requires the reduction of the gap distance, which is less constraining than other solutions. However, similar to cesium plasma, power is consumed to maintain the electric field of the gate, which reduces the overall efficiency [22]. The most straightforward way to reduce or nullify the space charge effect is to decrease the vacuum gap separating the cathode and anode. If the distance separating the two electrodes is less than 10 μm, the space charge effect can be effectively avoided [32] because there is neither sufficient space nor time for electrons to collide with each other and form a cloud [11]. Previously, interest in spacer-based TICs was constrained to academia due to the challenges of maintaining close spacing between two surfaces over large areas with extreme temperature differences [31]. Only recently were micron-gap TICs considered feasible and reliable options for energy converters, mostly due to advancements in nanofabrication technologies [22], [32]. Plasma-based converters are able to achieve power densities above 10 $W.cm^{-2}$ [38], but their efficiencies remain low (<15%) due to the plasma's energy consumption [11]. Including a gate between the electrodes has so far produced only low power densities (on the order of $mW.cm^{-2}$) [39], mainly due to electron shading by the gate [22]. However, spacer-based solutions (i.e., utilizing small pillars to physically separate the hot and cold sides of a device) have achieved a power density of 1.5 $W.cm^{-2}$ [40] and are less complex than plasma- or gate-based solutions, thus representing an interesting alternative to mitigate the space charge effect.

## 2.2. THERMOPHOTOVOLTAIC GENERATORS

Research and development of TPV devices began after that of thermionics. The first TPV device was fabricated by Henry H. Kolm in 1956 and consisted of a silicon solar cell irradiated by a Coleman camping lantern with a gas mantle [5], [41],[42]. However, Pierre Aigrain is typically credited with the invention of the TPV device [42], and his lecture series proposing direct heat-to-electricity conversion via thermophotovoltaics at MIT from 1960–61 was influential in motivating TPV research. Early TPV research focused on the development of low weight and portable electric generators for the military. The energy crisis in the mid-70s motivated research on solar powered applications for TPV, but the lack of high-quality photovoltaic cells stifled interest until the development of GaSb [43] and InGaAs cells in 1990 [44]. This spurred new developments, most notably a thermal-to-electric conversion efficiency of 23.6% in 2004 [45], which was only recently surpassed [17]. Several full systems were constructed up until the early 2000s, including the first commercial TPV device developed by JX Crystals [46]. In the following years the interest in TPV slowed despite the commencement of research on near-field TPVs (NF-TPV) [47], [48]. The current renewed interest in TPV devices can be attributed to the need for energy storage systems, and the development of thin-film InGaAs photovoltaic cells [49], as these cells can realize very high efficiencies at high temperatures [14]. It is important to note that current TPV systems require high operating temperatures to be economically viable.

Compared to traditional solar photovoltaics, TPV converters can have lower angular mismatch losses due to the close proximity of the emitter with the cell and spectral energy losses can be limited through photonic engineering [14]. However, their output power densities are still lower than that of thermoelectric devices [50] and fundamentally limited because radiative heat transfer from the emitter to the photovoltaic cell is constrained by the Blackbody limit [51]. In recent years, research has not only focused on improving the efficiency of photovoltaic cells, but also on methods to overcome the Blackbody limit; this would reduce the cost of electricity per watt, especially at lower emitter temperatures, allowing the use of TPVs in a wider array of applications such as waste heat recovery.

Increasing the power density of TPV devices is crucial to reducing the cost of electricity per watt. As previously indicated, one method of augmenting the power density is raising the emitter temperature [14]. However, other pathways exist, such as light pipes, in which an intermediary material is used between the emitter and the cell [52], and electroluminescent heat pumps, where a diode is used as an emitter [53]. Both approaches are susceptible to optical losses limiting their practical applications. Near-field thermophotovoltaics are another way to enhance the power density of TPV systems. In NF-TPV devices, the vacuum gap between the emitter and receiver is scaled down to the order of the thermal radiation wavelength [5], [54]. At such nanoscale distances, evanescent waves—electromagnetic waves generated by total internal reflection and that extend the distance of approximately one wavelength normal to the surface of a heat-emitting body [55]— can travel from the emitter to receiver, exciting atoms in the receiver and enabling heat transfer in a process called photon tunneling. If the thermal emission source and photovoltaic cell in a TPV device are separated by distances of the order of nanometers, both propagating and evanescent waves contribute to radiative heat transfer, which vastly enhances the output power density. This effect allows NF-TPV devices to output the same power densities as far-field devices at much lower emitter temperatures. This is of particular interest in systems featuring recovered waste heat as the heat source since temperatures generated are lower than other sources. The concept of NF-TPVs was first proposed in 2000 [47], and in 2001, DiMatteo *et al*. [48] observed that the short-circuit current of the PV cell increased when the space between the heat source and the cell was diminished. Since then, several theoretical studies have emerged on the concept of NF-TPVs [56], [57], [58], [59], while experimental works have focused on heat transfer measurements between two planar surfaces without a PV cell [60], [61], [62], [63], [64]. In 2018, Fiorino *et al*. [65] developed an NF-TPV device that achieved a 40-fold enhancement in output power density compared to a far-field device by implementing a positioner to create a gap less than 100 nm. Other works also demonstrated that output power density increases for gaps around 100 nm [66], [67], [68], [69]. The highest power density reported in a near-field configuration (0.5 – 0.75 $W.cm^{-2}$) was achieved using a piezoelectric activated tip as an emitter to realize gap distances below 100 nm [68], [69].

## 2.3. STATE OF THE ART OF SPACERS IN TPV AND TIC

Although the method of electricity production in TICs differs from that of NF-TPV devices, the performance of both technologies can be improved by shortening the distance between the emitter and the receiver. Several approaches exist to produce a nanometer or micrometer scale gap between an emitter and receiver: actuator-based [65], [70], MEMS based [71]; and spacer-based [47], [71], [72]. Actuator-based approaches are advantageous in experiments because they rely on positioner stages, enabling a large range of gap distances with the same setup, but they are difficult to implement at larger scales. Thus far, MEMS-based designs have been limited to small active areas compared to the volume of the device, which introduces heat losses [74]. Furthermore, MEMS-based systems consume energy to function and are subject to thermal stress. Spacer-based approaches, where a material is intercalated between the emitter and receiver to create a nanoscale gap, are advantageous because they allow for one-body devices without moving parts and are scalable to larger areas. These designs also rely on microfabrication techniques that are well established in the semiconductor industry, and with progress made in fabrication in recent years, spacers are becoming more and more feasible [54].

To reduce the space charge effect in thermionics, gaps should be of the order of a few microns, with the ideal distance depending on the temperature of the emitter [29]. For gaps smaller than 1 µm, the near-field effect becomes prevalent, enhancing the radiative heat transfer between the cathode and the anode. While this phenomenon is desirable for NF-TPVs, in TICs, this radiative heat flux competes with the electron flux and is an important source of losses [29]. For NF-TPV applications, gaps lower than 300 nm should be targeted to take full advantage of the near-field effect [75]. In both applications, the spacers share similar requirements: they should be able to withstand high temperatures without breaking, but they should also be designed to limit conduction through the spacer to prevent heat losses. Furthermore, the material of the spacer should have a high electrical resistivity to be electrically insolating and prevent short-circuiting of the device [22].

This work will first give a summary of existing works using spacers to enable micro- and nanoscale gaps before delving into the challenges spacers present, specifically addressing materials, heat losses and fabrication. For both TIC and NF-TPV, several articles did not focus on constructing a device that produces electricity. Most works targeting NF-TPV applications characterize near-field radiative heat transfer between two surfaces without converting heat into power. Research on TICs results in device fabrication more often than NF-TPVs, but several studies focus only on spacer characterization without thermionic conversion. For clarity in the manuscript, the distinction between articles pertaining to NF-TPVs and TICs has been made according to the authors' intended use, but methodology and findings could potentially be applied interchangeably to both applications.

### 2.3.1. Spacers used in thermionic devices

The first thermionic converter using spacers was created by Beggs [35]. In their device, three pieces of molybdenum foil were used to maintain a gap under 6 µm between the anode and cathode. At an emitter temperature of 1423 K, an output power of 1 $W.cm^{-2}$ was achieved with a reported efficiency of 4–5%. In 1993, a thermionic converter was fabricated by Fitzpatrick *et al.* [76] using alumina spacers to form a 9.5-µm gap. Very little information about the spacers was provided but an 11.6% lead efficiency was reported for a temperature of 1300 K. Experiments were conducted under a cesium pressure of 20 or 132 Pa, with the highest efficiency achieved for the lower cesium pressure. From 1999 to 2001, King *et al.* [77], [78] published a series of works demonstrating a TIC with silicon dioxide spacers sandwiched between a cathode consisting of a sapphire substrate and an anode comprised of a silicon dioxide substrate. A layer of a chromium-tungsten alloy and a BaO/SrO/CaO coating were applied to both electrodes to lower the work function. The center of the silicon dioxide anode was etched to form a gap with a minimum distance of 15 μm; the role of the spacer was played by the non-etched portion of the anode, which formed posts. The next TIC device featuring spacers would come more than ten years later, with Lee *et al.* [79] fabricating a micro-TIC, shown in Figure 2a), with a suspended emitter separated from the collector by silicon dioxide posts. In this work, the poly-SiC emitter coated with BaO/SrO/CaO/W was supported by suspension legs with a U-shaped cross section to increase out-of-plane rigidity and avoid contact between the cathode and anode during heating. Silicon dioxide posts served as support for the suspended emitter structure. This spacer design provided thermal insulation between the cathode and anode and demonstrated stability for several hours at temperatures ranging from 900 K to 1400 K. In 2014, the same group proposed an improved design, in which the emitter was coated with a thin tungsten layer to improve the adherence of the barium oxide. A 5-μm gap was realized, compared to the 10-μm gap previously reported [80]. Littau *et al.* [81] achieved a gap of approximately 5 μm using alumina microspheres to maintain the gap between a barium-coated tungsten cathode and a tungsten-coated silicon anode. The stated gap size was approximated as the beads' diameters had a large size dispersion. For example, a device with nominal bead diameters between 5 μm and 7 μm produced a thermionic emission I-V curve that better fit the I-V curve of an 11- μm-gap. The beads moved around while operating the device, which could cause issues with reliability, as noted by the authors. In 2014, Belbachir *et al.* [28] separated a SiC cathode and an anode formed by a thin platinum film by fabricating 56 square, silicon dioxide columns on the anode, achieving a 10-μm gap. This work focused heavily on the thermal characterization of the spacers. Notably, the thermal resistance of the spacers was evaluated in [28], this is a critical aspect of spacer design that is discussed in Section 4.2. Conduction losses through the $SiO_2$ columns were measured by characterizing the thermal resistance of the vacuum gap using calorimetry in an ambient atmosphere. First, the thermal resistances of the cathode and anode were assessed independently, followed by the thermal resistance of the full stack, which consisted of cathode/spacer/anode. At 100 ºC, the thermal resistance of the entire vacuum gap, including contact resistances between the cathode and spacers, and between the spacers and anode, was 2.4 $K.W^{-1}$. This value was used to approximate the conduction losses through the spacers at 830 ºC and was compared to the conduction losses calculated using Fourier's law and the bulk thermal conductivity of silicon dioxide. For the given experimental conditions, the thermal resistance measurements yielded conduction losses of 193 W, while calculations using the thermal conductivity of silicon dioxide resulted in 380 W of conduction losses. In 2020, Campbell *et al.* [40] proposed a removable spacer design made of a corrugated film with a hexagonal honeycomb pattern and a U-shaped cross-section to enhance strength and stiffness.

This design was based on a design by Nicaise *et al.* [30] (discussed in section 2.3.2.). In [40], alumina, an alumina-hafnia alloy, and alumina-hafnia-zirconia nanolaminate films were investigated as spacer materials, as the thin layers in the nanolaminate films increase phonon scattering. Although the article focused on the thermal characterization of the spacers and spacer materials, a TIC was also developed that achieved a 10 $A.cm^{-2}$ short-circuit current density for an emitter temperature of 1280 K. In 2021, Bellucci *et al.* [82] compared the output power densities of a device using zirconia spacers (see Figure 2b)) to form a 3-μm gap and a device employing actuator stages to form a 125-μm gap. For cathode temperatures below 1350 °C, the spacer-based device demonstrated a higher output power density compared to the actuator-based device. The cylindrical spacers were arranged in a ring pattern on the anode to minimize the contact surface between the emitter and the receiver. In 2018, Trucchi *et al.* [83] proposed a hybrid thermionic-thermoelectric generator using an alumina ring embedded in the collector to form a 100-μm gap in the thermionic portion of the device. In this hybrid device, concentrated solar radiation heated an absorber, which in turn heated a thermionic emitter. As the electrons produced by thermionic emission were collected by the anode, the heat transferred to the anode was used to heat a thermoelectric generator, producing power. A thermal-to-electric conversion efficiency of 6% was achieved, although the authors identified several ways to optimize the design.

### 2.3.2. Characterization of spacers for application in thermionics

A few articles focused solely on the characterization or fabrication of spacers, specifically for thermionics. In 2009, Yao *et al.* [74] proposed semicircular hollow cylinders formed by SU-8 photoresist to achieve a 10-μm-gap and measured their thermal conductivity at temperatures up to 350 K. Although the spacer design was intended for use in a thermionic micro-battery, the authors noted that the spacer material should be replaced for operating temperatures above 400 °C. In 2016, Belbachir *et al.* [84] reused the design of their previous experiments described in the section 2.3.1 to further investigate the vacuum gap's thermal resistance, specifically the impact of contact pressure on thermal resistance. The sum of the thermal contact resistances between the spacer and each surface increased as the width of the silicon dioxide spacers decreased. Furthermore, the thermal contact resistances between the spacers and electrodes accounted for roughly 80% of the total thermal resistance of the vacuum gap; thus, interfacial resistances could be critical in reducing parasitic conduction losses. In 2019, Nicaise *et al.* [30], fabricated an alumina spacer in the form of a corrugated film with hexagonal unit cells, as depicted in Figure 2c). This spacer film was advantageous because it was completely removable, which drastically enhanced the thermal contact resistance at the interfaces between the electrodes and spacers compared to spacers that are directly fabricated onto one electrode surface. The thermal resistance of the alumina spacer film was characterized in detail, and a small effective thermal conductivity of 5 $mW.m^{-1}.K^{-1}$ was achieved for a spacer size of 3 μm. This work was further built up on in 2020 with a similar design, but small protrusions referred to as "bumps" were placed at the hexagon intersections to augment the interfacial contact resistance and multilayered materials were used as discussed in section 2.3.1.[30]. Although the gap distance in ranged from 1.8 μm to 2.2 μm, the authors concluded that this design potentially could be scaled down to 100 nm to create nanogaps useful for NF-TPV applications.

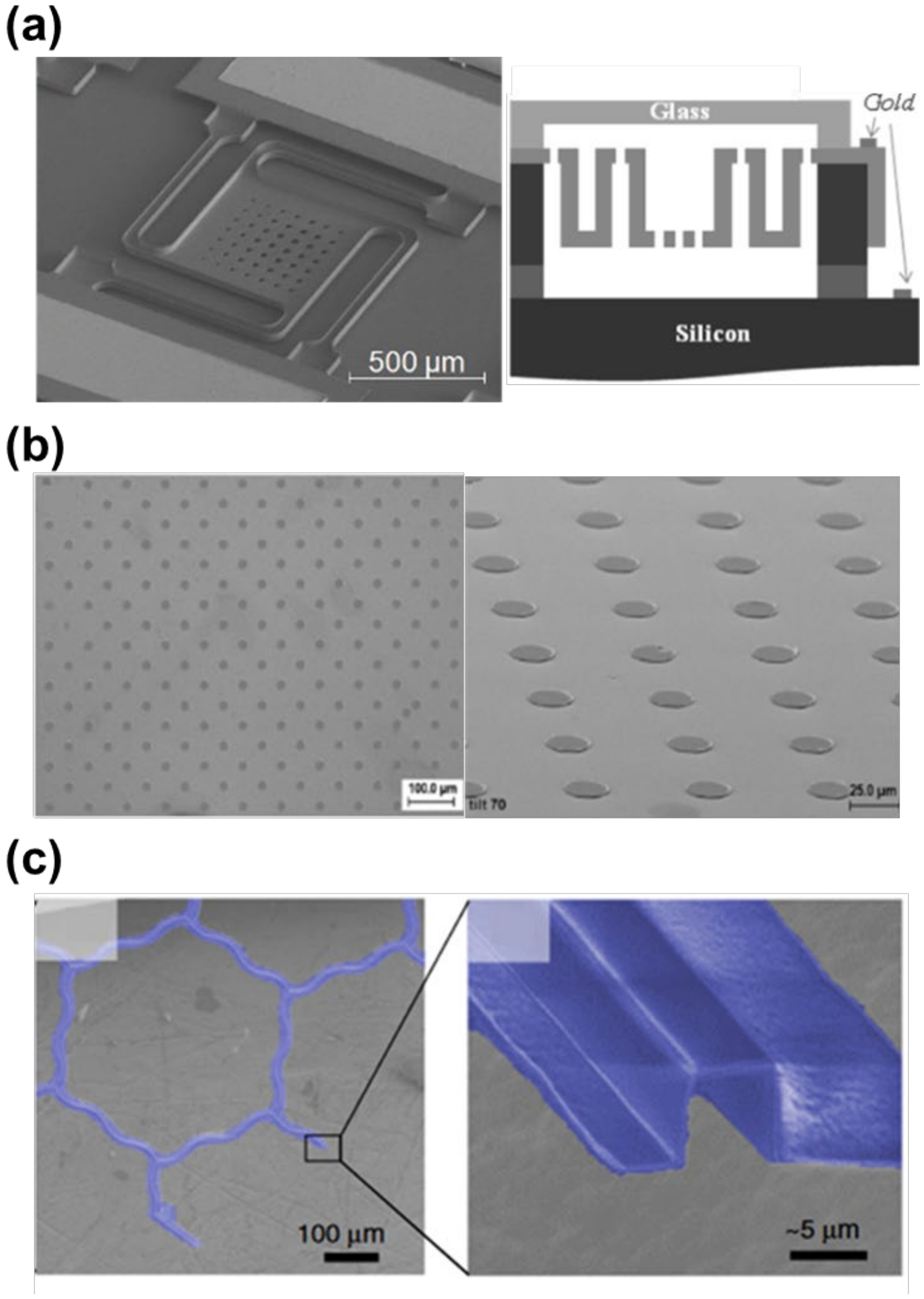

*Figure 2 – Illustrations of different spacer designs for TICs: a) SEM image (top) of a micro-TEC, and a cross-sectional diagram (bottom) of the device,as presented in [79] b) SEM images of the $ZrO_2$ spacers used in [82] c) Corrugated, hexagonal spacer as presented in [30].*

### 2.3.3. Spacers used in near-field thermophotovoltaic devices

To our knowledge, only five articles have resulted in actual NF-TPV devices featuring spacers. The first spacer-based NF-TPV device was proposed by DiMatteo *et al.* [48] in 2001. Cylindrical, silicon dioxide spacers were fabricated onto a silicon heating chip, which acted as the emitter, to achieve a 1-μm gap. A piezoelectric actuator was also employed to induce oscillations of the emitter and the gap and measure its impact on the short-circuit current of the TPV cell. The short-circuit current was shown to increase as gap size decreased, suggesting the impact of the gap size on the number of photons being absorbed in the TPV cell. DiMatteo *et al.* [85] made further progress in 2004 by achieving a higher emitter temperature, smaller gap distances and a larger emitter area. Also, compressible tubular spacers were placed within pits etched on the emitter, as opposed to the solid cylinders in [48]. Calculations indicated that the compressible spacers' hollow nature and increased length created by the pits reduced parasitic heat conduction by nearly 93%, from 3.78 W.cm$^{-2}$ in [48] to 0.27 W.cm$^{-2}$ in [85]. In both articles, the gap distance was determined by measuring the capacitance between the emitter and PV cell, whereas most subsequent works measured the gap by interferometry or simply assumed that the gap was equivalent to the size of the spacers. Spacers were not used in NF-TPV devices for many years until 2019, when Inoue *et al.* [66] developed a one-chip design integrating the emitter, spacers and receiver into one compact device. This device featured a silicon emitter that was suspended over an intermediate silicon substrate and a thin-film InGaAs cell by a supporting beam. Although the gap was created by the suspension of the emitter above the intermediate substrate, the full intermediate silicon layer was considered the spacer in this design. This layer contained pits with a single supporting pillar that enabled the use of reflection spectra measurements to assess the gap distance. Additionally, the intermediate layer served as a transmission medium towards the PV cell. Although thermal deformation of the emitter was observed, resulting in a larger gap size than intended, a gap size of 140 nm was still achieved, as well as a large temperature difference (>700 K) between the emitter and the PV cell. A 10-fold enhancement in the short-circuit current and output power were observed in the near-field device compared to a far-field device with a 1160-nm gap. The same group further improved the design in 2021 by adding supporting beam at all four corners of the emitter [86], as shown in Figure 3a), instead of a single beam. The emitter thickness was also increased from 2 μm in [66] to 20 μm in [86] to augment the density of photonic states inside the emitter. The single-beam device featured in [66] was not rigid enough to prevent contact between the emitter and intermediate substrate, but the more robust device in [86] output a power density of 192 mW.cm$^{-2}$ at emitter temperature of 1192 K, resulting in a system conversion efficiency of 0.7%. It should be noted that the active heat transfer areas in [66] and [86] were very small (1 $mm^2$). However, Selvidge *et al.* [72] recently introduced a self-supported emitter-cell device with an area of 0.28 $cm^2$ and a 150-nm gap, as shown in Figure 3b). The spacers in this device were four long, GaAs posts formed through etching of GaInP/GaAs emitter. The device generated 4.4 $mW.cm^{-2}$ at 460 ºC and demonstrated a short-circuit current that far exceeded the photocurrent limit, confirming super-Planckian behavior.

### 2.3.4. Characterization of spacers for applications in thermophotovoltaics

Most researchers' efforts involving spacers for NF-TPV applications have focused on plane-to-plane experiments without PV cells, to experimentally demonstrate the near-field effects that enable radiative heat transfer surpassing the Blackbody limit. In 2008, Hu *et al.* [60] investigated the radiative heat flux between two glass plates with a gap maintained by polystyrene spheres with 1-µm diameters, as depicted in Figure 3c). The measurements were conducted at room temperature, and the experimental setup achieved a temperature difference between the emitter and receiver of 83 ºC. This configuration produced a radiative heat flux that exceeded the Blackbody limit by 35% across the entire range of emitter temperatures (~310–340 K). In 2015, Ito *et al.* [61] investigated the radiative heat transfer between two planar silicon dioxide substrates separated by spacers of the same material in the form of truncated pyramids, as illustrated in Figure 3d). The spacers were fabricated on both the emitter and receiver surfaces to achieve a 500-nm gap. The thermal resistance of the pyramidal structures was also modeled to approximate conduction losses through the spacers. For all of the spacer lengths investigated (0.5 µm, 1 µm and 2 µm), the contact resistance between the spacers and the $SiO_2$ substrates far outweighed the resistance of the bulk of the spacer, confirming the potential of contact resistance in mitigating conduction losses. In 2017, Ito *et al.* [87] reduced the gap distance to 370 nm and replaced the receiver with a silicon dioxide substrate coated with vanadium oxide, taking advantage of the metal–insulator transition of tungsten-doped vanadium oxide to modulate the radiative heat flux. In 2016, Watjen *et al.* [62] measured the radiative heat transfer between two silicon plates with areas of 1 $cm^2$ separated by a minimal gap of 200 nm, which was created by silicon dioxide columns. The gap distance was determined by infrared reflectance measurements, and the radiative heat transfer was approximated by subtracting the estimated conduction losses through the spacers from the total heat flux, which was measured by calorimetry. During the same year, Bernardi *et al.* [63] also measured the radiative heat transfer between two silicon plates, but their design included both SU-8 and silicon dioxide posts to form a 150-nm vacuum gap. The silicon dioxide posts had a height of 150 nm; they were located in the center of the device and served as a stopping point for the emitter when pressure was applied, as shown in Figure 3e). The 3.5-μm-tall SU-8 spacers were placed at the edge of the device and were only used to maintain the emitter suspended in far-field conditions. This design allowed for a tunable gap distance, but it introduced some uncertainties in properly defining the active heat transfer area and by consequence in quantifying properly the heat transfer. For a temperature difference of ~116 K, the maximum radiative heat flux between the two 25-mm$^2$ silicon plates was 8.4 times larger than the Blackbody limit.

In 2017, Lang *et al.* [64] used silicon dioxide nanospheres to maintain a 150-nm gap between two glass plates. While the vast majority of researchers measured the steady-state heat transfer across the vacuum gap, the measurements in [64] represent the transient heat transfer. Despite a very minor temperature gradient of 7 K, the observed near-field heat flux was 16 times larger than the Blackbody limit. In 2018, Yang *et al.* [88] coated two planar silicon substrates with graphene to explore whether graphene enhanced or diminished the radiative heat transfer between the surfaces in a design illustrated in Figure 3f). In this setup, AZ photoresist was utilized to fabricate posts and achieve a gap of 430 nm. A larger heat transfer coefficient was observed when both the emitter and receiver were coated in graphene due to plasmonic mode coupling, demonstrating the utility of graphene in enhancing near-field radiative heat transfer. One year later, DeSutter *et al.* [73] achieved gaps ranging from 110 nm to 900 nm using SU-8 pillars between two *p*-doped silicon plates. The cylindrical pillars were embedded in 4.5-μm-deep pits that were etched into the emitter surface, as depicted in Figure 3g). These pits allowed for pillars that were much longer than the gap distance, which increased the thermal resistance of the spacers without compromising the near-field enhancement in thermal radiation. For a device with a 110-nm gap, conduction losses declined by a factor of ~42 when the 4.5-μm-deep pits were incorporated into the device. Temperature differences up to 100 K between the emitter and receiver were investigated, but the design was inherently limited to lower operating temperatures due to SU-8's instability at temperatures exceeding 450 K. In 2018, the same group compared the radiative heat transfer for two device configurations: the first featured two *p*-doped Si surfaces, and the second was an SiC emitter and a receiver consisting of *p*-doped silicon. In both configurations, the emitter and receiver were separated by a 150-nm gap formed by silicon dioxide nanopillars [89]. The radiative heat transfer coefficient between the SiC-Si pair was slightly larger than its Si-Si counterpart, but the SiC-Si pair exhibited more monochromatic behavior due to surface polariton coupling between the dissimilar materials. In both [73] and [89], the active heat transfer areas were 25 mm$^2$. In 2020, SU-8 was utilized again as a spacer material by Ying *et al.* [90] to create a 190-nm gap between two *p*-doped silicon surfaces. The authors noted that SU-8 can be used to bond two substrates together to create near-field radiative thermal devices. Polystyrene spheres were employed as spacers by Sabbaghi *et al.* [91] due to the material's low thermal conductivity. The device was characterized by a 215-nm gap between two doped silicon surfaces coated with aluminum films between 13 nm and 80 nm in thickness. Thinner aluminum coatings corresponded to higher near-field radiative heat fluxes due to non-resonant electromagnetic coupling within the gap, known as the near-field effect, as well as resonant coupling within the thin Al layer, or the thin-film effect. The uncertainty in the vacuum gap distance was higher in this work $\left(215^{+55}_{-50}\ \text{nm}\right)$ compared to other works. However, unlike other publications using spherical spacers, Sabbaghi *et al.* [91] implemented the Hertz model to calculate the contact area between the polystyrene spheres and silicon surfaces when estimating conduction losses through the spheres. This work, published in 2020, was the last to use spheres as a spacer.

In 2021, Shi *et al.* [92] investigated the radiative heat transfer between two surfaces made of five layers of a graphene/SU-8 stack deposited on silicon dioxide. A gap of 55 nm was maintained through SU-8 pillars, as shown in Figure 3h). The Fermi level of the graphene was tuned by applying a voltage bias, revealing a strong enhancement of the radiative heat flux as a function of increasing Fermi level. The same team followed this work with another set of experiments, but this time, the graphene's Fermi level was tuned even further to couple and decouple surface plasmon polaritons across an 81-nm gap separated by SU-8 pillars. This tuning enabled modulation of the radiative heat transfer between the two surfaces [93]. Furthermore, Shi *et al.* [93] noted that this method should be feasible for other Van der Waals' heterostructures with the appropriate resonance modes. Graphene was also used by Lu *et al.* [94] in 2022 to demonstrate the enhancement of radiative heat transfer resulting from coupled surface plasmon polaritons and hyperbolic phonon polaritons. In this work, emitters and receivers composed of intrinsic silicon substrates with identical coatings were separated by a 400-nm gap created by four cylindrical AZ 5214 spacers. Two coatings were investigated: a graphene and hexagonal boron nitride (hBN) heterostructure, and a graphene/hBN/graphene sandwich. For a temperature difference of 40 K, the graphene/hBN heterostructures produced a radiative heat flux three times larger than the Blackbody limit, while the graphene/hBN/graphene multilayers resulted in a six-fold enhancement. The six-fold increase was attributed to strong coupling of surface plasmon polaritons and hyperbolic modes. In 2024, the same group similarly explored the impact of different emitters on the radiative heat transfer to a GaAs absorber [95]. Layers of indium tin oxide (ITO) or tungsten/silicon dioxide multilayers were added to a tungsten-coated silicon substrate. Again, the objective was to investigate the impact of surface plasmon polaritons and hyperbolic modes on radiative heat transfer. The gap between the hot and cold surfaces was maintained by five silicon dioxide pillars. When the gap distance was 400 nm, the experiments showed that the emitter featuring tungsten/silicon dioxide multilayers produced a radiative heat flux was 4.5 times larger than that of a far-field device (gap distance of 10 μm) due to coupled surface plasmon polaritons at the interfaces between metal and dielectric materials. Furthermore, the boost in radiative heat transfer produced by the multilayers was independent of the vacuum gap size, while the enhancement due to the ITO-coated emitter increased as gap size decreased. Li *et al*. [96] also investigated the impact of a film on near-field heat transfer by comparing the heat transfer between silicon substrates both with and without a 250-nm-thick silicon dioxide film. The 302-nm gap was maintained by an array of silicon dioxide posts. The addition of the silicon dioxide film increased the near-field radiative heat flux by a factor of ~10 when compared to the silicon substrate without a film. The impact of a graphene film on near-field radiative heat transfer was again investigated by Habibzadeh *et al*. [97]. Two scenarios were considered: 1) radiative heat transfer between two SiC surfaces; and 2) radiative heat transfer between lithium fluoride (LiF) and SiC surfaces. In both scenarios, cylindrical SU-8 posts were implemented to create a 120-nm gap. Unfortunately, the authors noted a large distribution in the spacer length and, consequently, high uncertainty in the actual gap distance. The radiative heat transfer was greater for similar materials than dissimilar materials, but an increase in the radiative heat transfer flux was observed after adding a layer of graphene to the LiF emitter due to the interplay of the surface plasmon polaritons of graphene with the surface phonon polaritons of the dielectric.

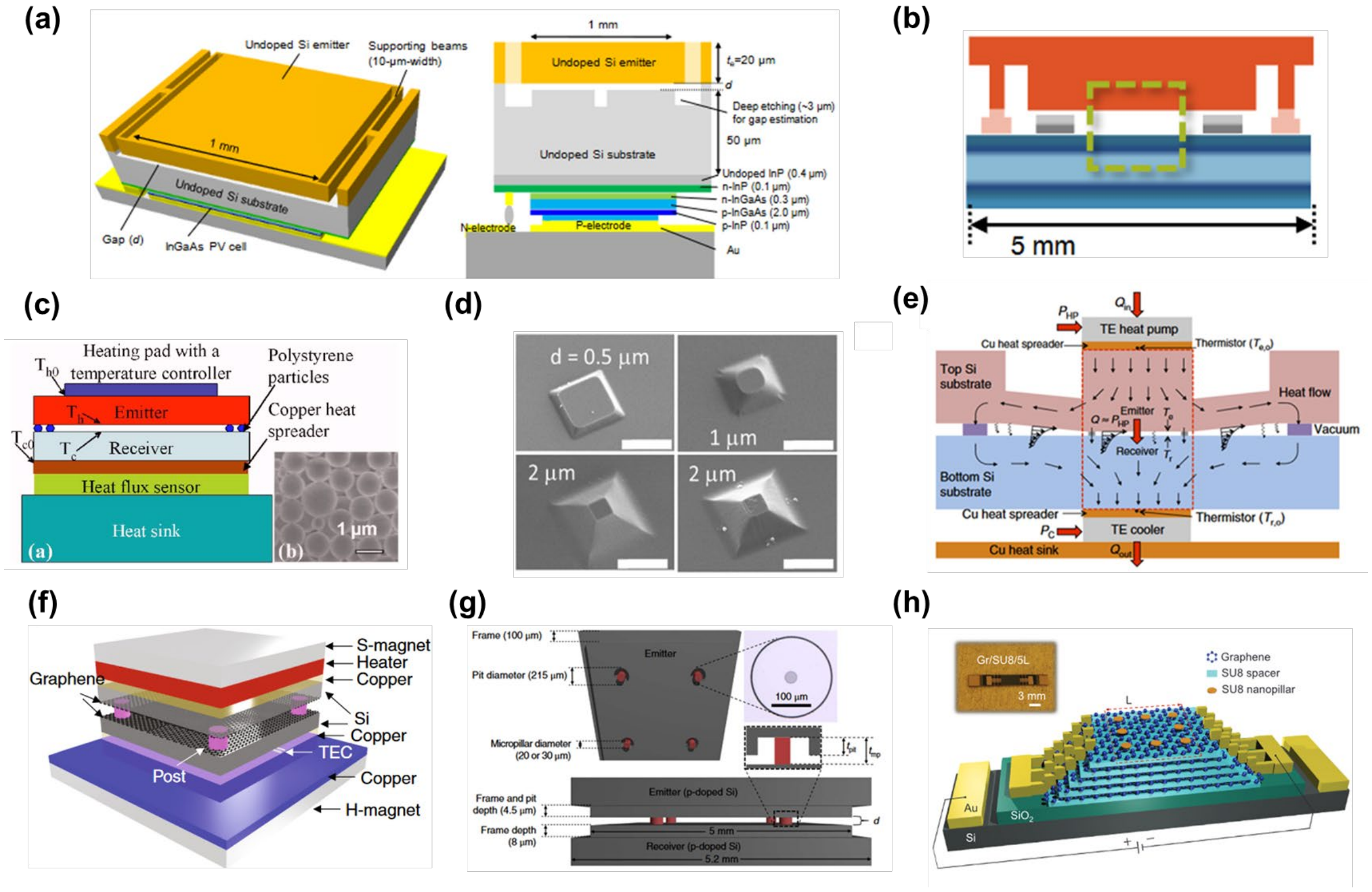

*Figure 3 – Illustrations of the different spacers strategies: a) A one-chip NF-TPV device as presented in [86] b) Design used by Selvidge et al. in [72] c) Schematic of the device proposed in [60] (left), and SEM images of the spherical polystyrene spacers that formed the vacuum gap d) SEM images of the truncated pyramids of various heights that were fabricated in [61] e) Device design used in [63] f) Drawing of the design in [88] g) Spacers embedded in the emitter as presented in [73] h) Design used in [92].*

## 3. DISCUSSION

This section provides an overview of the main parameters and experimental conditions of the works discussed in Section 2 to determine general trends in experiments regarding micro- and nano-gaps for NF-TPV and TIC applications. Table 1 summarizes several important parameters for characterizing NF-TPVs and TICs, including the materials of the emitter, receiver and spacer, the shape and distribution of the spacer, the minimum gap distance, the maximum temperature of the emitter, as well as the difference in temperature between the emitter and the receiver (ΔT) and the vacuum level at which the experiments were performed.

Figure 4 summarizes the preceding articles by illustrating the number of articles corresponding to a given spacer material for both TICs and TPVs. Articles that resulted in an energy conversion device and articles focused on characterizing the spacers' thermal properties or radiative heat transfer are separated by color.

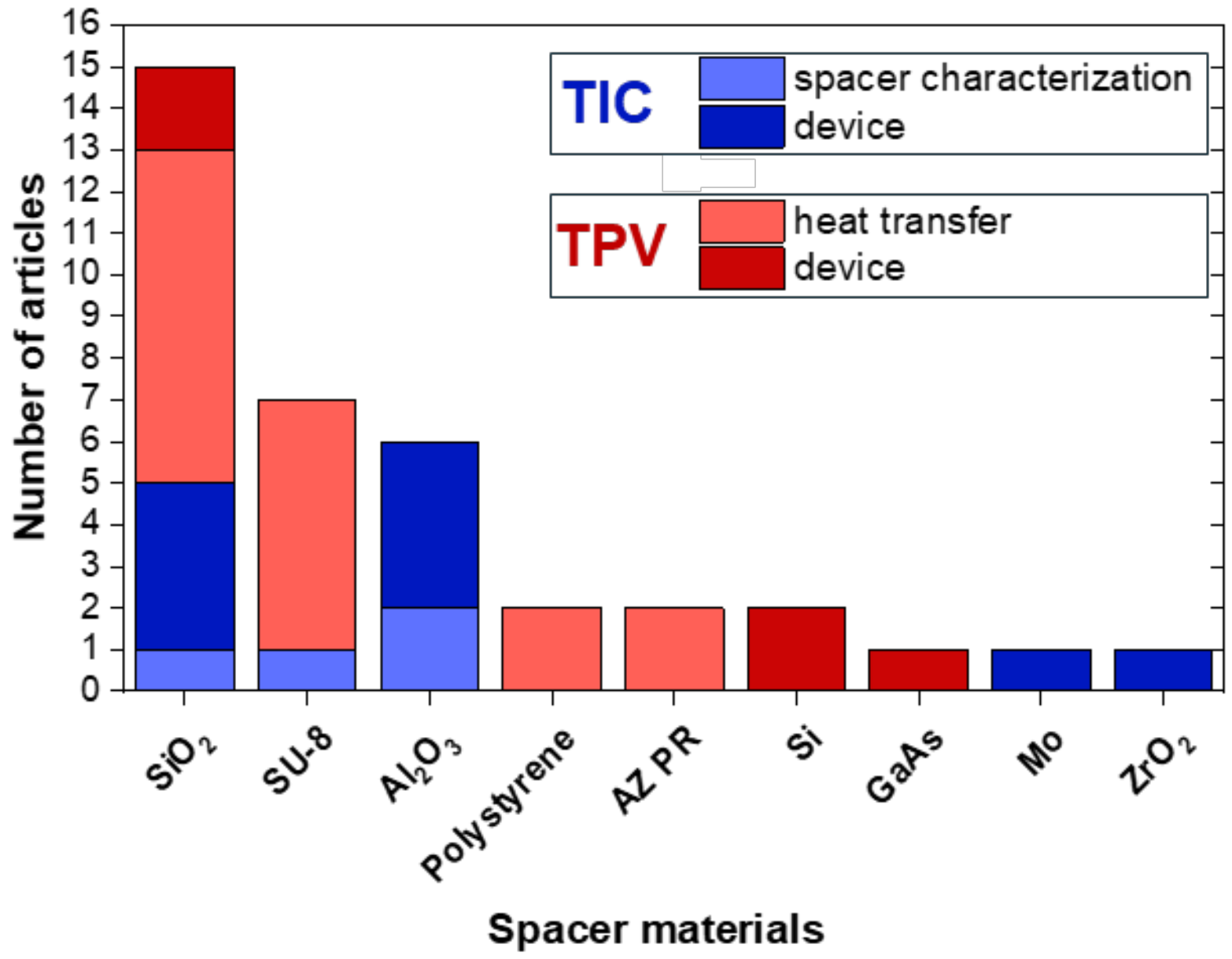

*Figure 4 - Number of articles using different spacer materials for NF-TPVs or TICs according to their application. Plane-plane near-field radiative heat transfer measurements and spacer characterization for TICs are denoted by light red and light blue, respectively. Actual NF-TPV devices are represented by dark red, while TIC devices are represented by dark blue.*

Most articles focused on heat transfer experiments intended for NF-TPV applications using plane-to-plane geometry. In the case of NF-TPVs, only five devices using spacers were fabricated. For TICs, the situation is reversed, as most of the published works demonstrated actual devices and very few articles focused solely on spacer characterization and fabrication. Figure 4 reveals that silicon dioxide is by far the most common spacer material, followed by SU-8 photoresist and alumina. Other organic materials have also been employed, such as AZ photoresist and polystyrene. The prominence of organic materials can be partially explained by their low thermal conductivities, as evident by the values in Table 2; this minimizes thermal conduction losses through spacers. In particular, SU-8, AZ Photoresist, and Polystyrene all have very low thermal conductivities of 0.2 and 0.18 $W.m^{-1}.K^{-1}$. Organic materials are not suitable for operation at the targeted temperatures for NF-TPVs and TICs. Out of all the materials listed in Table 2, only ceramics such as silicon dioxide, alumina and zirconia can be used at high temperatures. However, silicon dioxide and alumina have high thermal conductivities (1.3 $W.m^{-1}.K^{-1}$ and 2 $W.m^{-1}.K^{-1}$, respectively). Alumina is more thermally conductive than silicon dioxide, which may explain the prevalence of $SiO_2$ in NF-TPV and TIC experiments. Silicon dioxide is also advantageous due to its ease of patterning through existing, well-known techniques. Zirconia has a much lower thermal conductivity than alumina or silicon dioxide, and zirconia films have demonstrated higher electrical resistivity than alumina at operating temperatures of 1000°C [82]. However, it has only been employed once as a spacer material in a TIC device, and its performance in NF-TPV devices has never been evaluated. It should be noted that the thermal conductivity values presented in Table 2 are subject to change with pressure and temperature. Moreover at nanoscale, some material characteristic such as the presence of defects or the porosity can have an impact on the thermal conductivity [98], [99], [100]. These variations in thermal conductivity were not considered in most articles, which has introduced uncertainty in parasitic conduction loss calculations through the spacers.

*Table 2 – Thermal and electrical properties of the most common spacer materials in the literature.*

| **Material** | **Maximal temperature of use [K]** | **Thermal conductivity [W.m$^{-1}$.K$^{-1}$]** | **Electrical resistivity [Ω.cm]** | **Compressive strength [MPa]** |
|---|---|---|---|---|
| Silicon dioxide | 1986 [101] | 1.4 [102] | ~ $10^{11}$-$10^{16}$ [103] | 1100 [104] |
| SU-8 | 450 [105] | 0.2 [106] | ~ $10^{16}$ [101] | 108 [107] |
| Alumina | 2345 [108] | 2 [109] | ~ $10^{14}$ [101] | 2500 [104] |
| Polystyrene | 513 [110] | 0.18 [111] | ~ $10^{18}$ [101] | 0.750 [104] |
| AZ Photoresist | 420 [112] | 0.18 [113] | | |
| Silicon | 1687 [101] | 142 [114] | ~ $10^{5}$[115] | 120 [101] |
| Zirconia | 2988 [101] | 0.06 [116] | ~ $10^{10}$ [101] | 2200 [104] |
| Molybdenum | 2896 [101] | 137 [117] | ~ $10^{-6}$ [101] | 400 [101] |
| GaAs | 1513 [101] | 55 [118] | ~ $10^{7}$ [101] | |

Figure 5b) illustrates the frequency of spacer materials as a function of maximum emitter temperature. Most experiments were conducted at relatively low emitter temperatures (< 500 K), which is far from the target emitter temperature for NF-TPV and TIC applications. Most notably, organic materials have been implemented in a large proportion of spacer-based research experiments, with 11 articles accumulated between polystyrene, AZ photoresist, and SU-8. However, as discussed above, these organic materials are only practical for lower-temperature experiments because they would either become unstable or deteriorate completely at higher temperatures. SU-8 is the most common largely because of its ease of patterning through conventional photolithography techniques, but it is not viable in NF-TPVs or TICs.

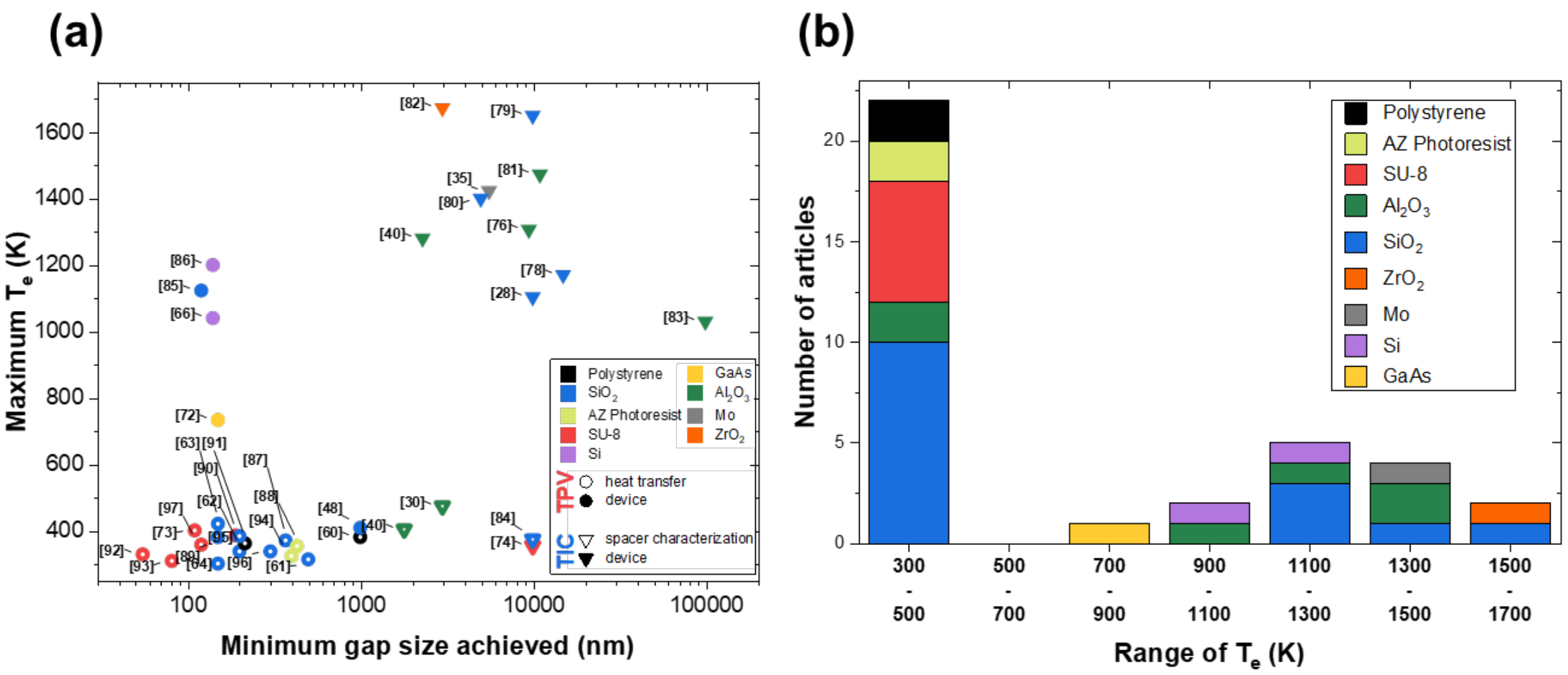

*Figure 5 – a) Maximum temperature of the emitter $T_e$ [K] corresponding to the different gap distances achieved in every article. Triangles represent experiments targeting TICs, with filled markers representing devices and hollow markers being experiments. Circles represent NF-TPV experiments with a similar distinction between devices and experiments. b) Number of articles that employed various spacer materials as a function of the range of emitter temperatures [K].*

By contrast, ceramic materials have been utilized at higher temperatures, especially silicon dioxide and alumina, which have both been used more than once in TICs with maximum emitter temperatures above 1300 K. Thus, these materials are much more suitable for thermionics and near-field thermophotovoltaics. Furthermore, ceramic materials have much higher compressive strengths than organic materials, metals or metalloids, as demonstrated by the values in Table 2. A high compressive strength is critical for spacer materials, as a force is often applied to the upper emitter surface to correct any bowing of the emitter or receiver, which aids in maintaining a uniform and parallel vacuum gap. However, this applied force can cause deformation or breakage of the spacer [22], [61], [82]. Therefore, the mechanical stability of the spacer materials under applied loads should be analyzed when selecting a spacer material.

At the device level, the area is a parameter of interest because spacers appear to be the most scalable solution compared to MEMS and positioners. Figure 6 illustrates the area of the surface hosting the spacer, either the receiver or the emitter, depending on the gap distance, for each article referenced in Table 1. Thermionic devices, represented by the points at gap distances above 1000 nm, achieved a bigger area (~1 $cm^{-2}$) than NF-TPV devices. This can be attributed to the fact that fabrication is less challenging at those larger distances and that it is easier to maintain parallel surfaces for larger gap distances, which is not the case for the distances of interest in NF-TPV devices. For TIC the biggest area for a device was 3.14 $cm^{-2}$ and was achieved by Campbell *et al.* [40] for an emitter temperature of 1280 K. The highest emitter temperature for a TIC device was1673 K, achieved by Bellucci *et al*. [82], but the area of the device was limited to 0.4 cm$^{-2}$. In the case of NF-TPV devices, the biggest area was realized by Selvidge *et al.* [72] with an area of 0.28 cm$^{-2}$, this experiment targeted the relatively low emitter temperature of ~700 K. Nonetheless it represents a considerable improvement compared to other devices, most notably the device by Inoue et al. [86] that achieved the highest emitter temperature (1200 K) but for a device area of 0.01 cm$^{-2}$. In addition, the design proposed by Inoue *et al.* appeared difficult to scale up due to the emitter potentially touching the intermediate spacer. Another notable NF-TPV device is the one fabricated by DiMatteo *et al.* [85] where a 0.2 cm$^{-2}$ emitter area was achieved with the caveat that this area was represented an array of four heater chips (of individual area of 0.05 cm$^{-2}$) instead of one emitter. Regarding non-devices, larger areas have been accomplished across the range of gap distances with several experiments achieving areas over 1 cm$^{-2}$. However, for gap distances below 100 nm, only small-scale areas (0.09 cm$^{-2}$) have been reported [92], [93], and only for low emitter temperatures, highlighting remaining challenges in scaling up the area for sub-100 nm gaps.

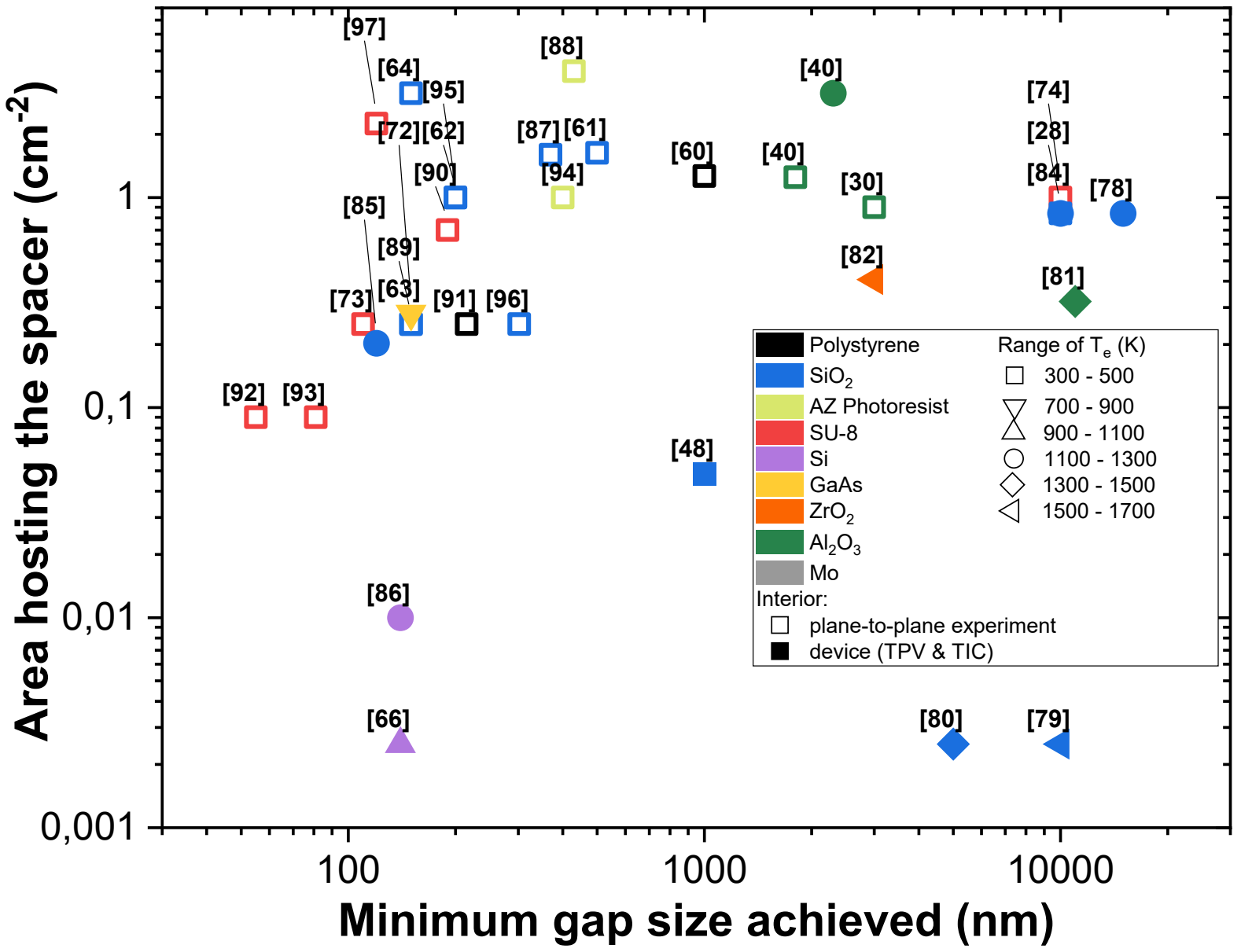

*Figure 6 – Area of the surface hosting the spacer (in cm$^{-2}$) plotted against the minimum gap achieved (nm). Filled markers indicate a device (for both technologies) while the shape of the marker indicates the range of the emitter temperature (in K).*

In both technologies, the flux of interest, which is electrons for TICs and photons for NF-TPVs, is not the only heat flux present. Convective heat transfer is suppressed because experiments are conducted in a vacuum, but the introduction of spacers between the emitter and receiver facilitates conductive heat transfer. This conduction competes with the useful energy transfer and represents an unavoidable source of losses for the system because the heat transmitted this way doesn't participate in the generation of electrical power. Figure 7a) shows the conductive heat transfer coefficient as a function of the minimum gap achieved for the articles that gave enough information to calculate the values [60], [62], [63], [64], [73], [88], [89], [90], [91], [92], [93], [94], [95], [96], [97]. The conductive heat transfer coefficient $h_{cond}$ is obtained from the conduction losses $Q_{cond}$ and the temperature gradient between emitter and receiver by the relation: $h_{cond}=Q_{cond}/\Delta T$. Minimizing this coefficient is key to avoiding parasitic heat conduction. All the values of $Q_{cond}$ used to calculate $h_{cond}$ were calculated by the authors with a Fourier's law (see Equation (2) in Section 4.2.), with the exception of DeSutter *et al.* [73] and Ito *et al.* [61], [87] both of whom employed finite elements methods (FEM) to estimate the losses. The results illustrated in Figure 7a) indicate that the conductive heat transfer coefficient increases as the gap distance decreases. The lowest value of $h_{cond}$ (2.3 $W.m^{-2}.K^{-1}$) was achieved by DeSutter *et al.* [73] for a 100-nm gap formed by embedded posts. For larger-scale gaps, the lowest conductive heat transfer coefficient was 0.24 $W.m^{-2}.K^{-1}$ and was attained by Hu *et al.* [60] using polystyrene spheres to create a 1-µm gap.

Figure 7a) also indicates the radiative heat transfer coefficients for different emitter temperatures, calculated by the Stefan-Boltzmann law. For high emitter temperatures ($T_e$>1200 K), radiative heat transfer exceeded conductive heat transfer for most spacer-based setups, except for the experiments in which the gap was less than 100 nm [92], [93]. However, at those gap distances, near-field effects enhanced the radiative heat transfer by several folds, resulting in a vacuum gap that was still dominated by radiative heat transfer, not parasitic conduction through the spacers. As expected, materials with lower thermal conductivity, such as SU-8 and polystyrene, produced a smaller conductive heat transfer coefficient than materials with higher conductivities for similar gap sizes.

To mitigate conduction losses through spacers, four distinct strategies can be identified, namely:

- Reducing the surface coverage, which is defined as the ratio of area of the emitter or receiver occupied by spacers (see Table 1). As shown in Figure 7b, the surface coverage is reduced as the gap distance is decreased in an attempt to counterbalance the higher thermal conduction losses. However, upon examining the percentage of total heat transfer comprised by conduction losses, shown in Figure 7c), this approach does not appear to be successful, as the reduction of surface coverage did not necessarily lead to lower conduction losses.
- Embedding the spacers into the emitting or receiving surface. Another approach applied by a few authors, is to use the length of the spacer to increase the path of heat conduction through the spacers, consequently reducing conductive heat flux.

DeSutter *et al.* [73] fabricated spacers in 4.5-μm-deep pits (as seen in Figure 3g) to lengthen the conduction pathway while still achieving gap distance nearing 100 nm. Conduction losses were estimated to represent about 4% of the total heat transfer which lead to a conductive heat transfer coefficient of about 2 $W.m^{-2}.K^{-1}$. This is the lowest value of $h_{cond}$ achieved in this gap range. DeSutter *et al.* [73] stated that conduction losses would constitute 45% of total heat transfer if the same device was fabricated without pits. Selvidge *et al.* [72] employed a similar approach to form a 120-nm-gap. In this case, conduction losses were estimated as 19–26% of the total heat flux.

- Optimizing the shape and material of the spacer. Some examples include implementing hollow or tubular spacers, as proposed in [74], [85], or fabricating pyramids with a truncated pyramid shape where the top area can be optimized to increase the thermal resistance [61]. Ito *et al.* designed truncated pyramidal spacers from $SiO_2$ and were able to realize a 370-nm gap and a conductive heat transfer coefficient of 6 $W.m^{-2}.K^{-1}$.This value was the lowest reported for silicon dioxide spacers in this range of gap sizes, but Figure 7c) indicates that trapezoidal spacers were only somewhat successful in mitigating conduction losses, as 34% of the total heat transfer was still conductive. Employing a material with a low thermal conductivity can also be a solution, as illustrated in Figure 7c), the lowest conduction losses were achieved using SU-8 and polystyrene.
- Using unattached spacers. Designs in which spacers were not attached to either surface increase interfaces, and this can reduce conduction losses considerably [30], [40], [64]. This strategy will be discussed in detail in Section 4.2.

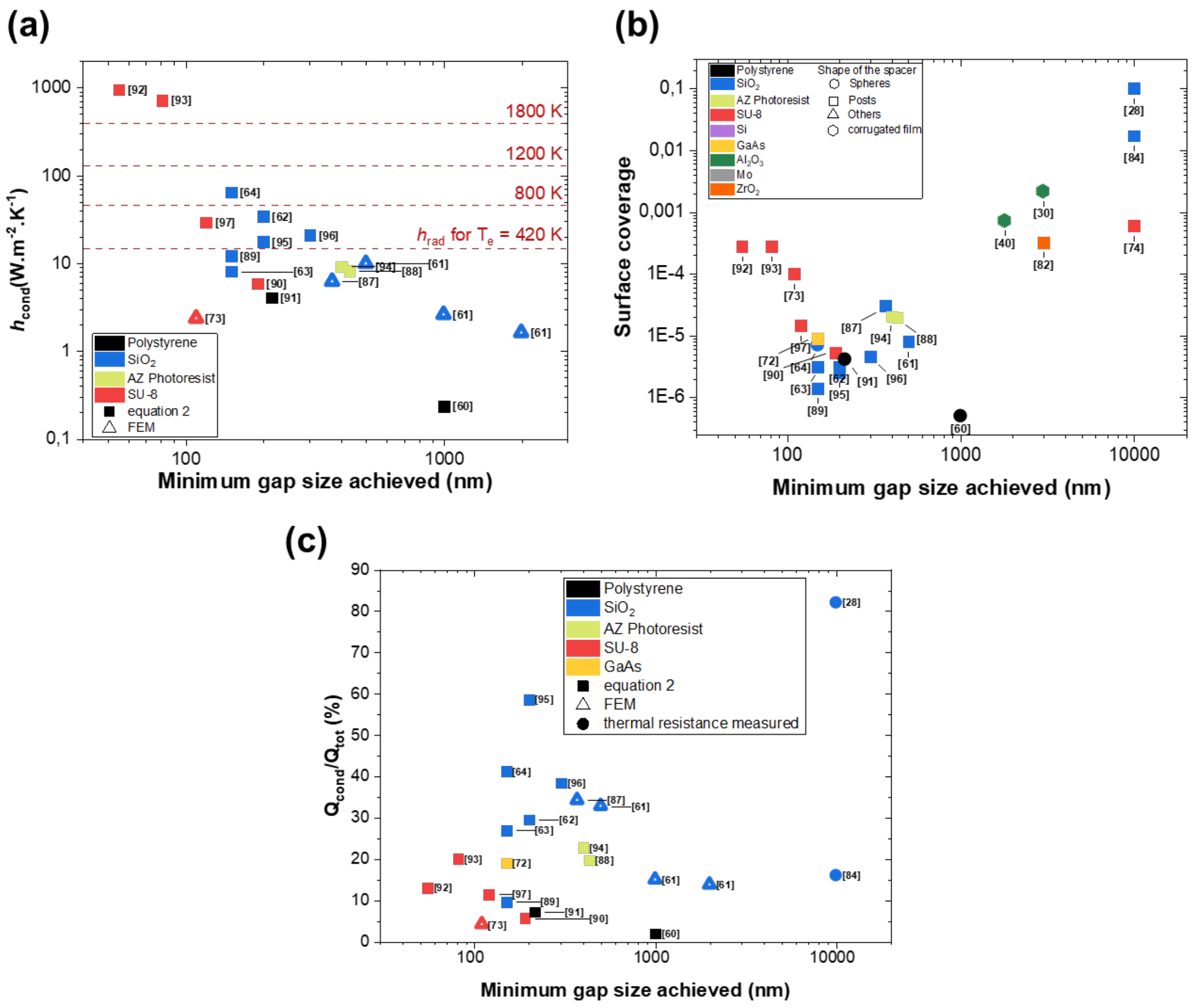

*Figure 7 – a) Conductive heat transfer coefficient versus minimum gap distance. The red dashed lines are the estimated radiative heat transfer coefficient for the maximum emitter temperature in this group of experiments (420 K). The estimated $h_{rad}$ values corresponding to higher emitter temperatures (800, 1200, 1800 K) are provided as an example. The pool of data here is limited due to the lack of data available in published works. b) Surface coverage of the spacer structures versus the minimum gap achieved. c) Percentage of losses due to conduction through the spacers at various gap sizes.*

Finally, it is worth noting that most articles used Fourier's law to determine the conduction losses with only a few relying on FEM or thermal resistance characterization of the vacuum gap. Utilizing Fourier's law to evaluate conduction losses neglects interfacial contact resistances between the spacers and hot/cold surfaces, which results in overestimated conduction losses. The importance of thermal contact resistance in determining conduction losses will be discussed in detail in Section 4.2.

## 4. EXPERIMENTAL UNCERTAINTIES

Several authors discussed the difficulties in determining the contribution of radiative heat transfer in an accurate manner. Typically, the total heat flux from the emitter to the receiver is the only heat flux that is measured experimentally; thus, the radiative and conductive heat fluxes must be isolated from the total measured value. Measuring the vacuum gap distance precisely has also challenged authors, and such gap measurements are crucial in verifying experimental results with theoretical predictions.

### 4.1. COEXISTENCE OF HEAT TRANSFER MECHANISMS

The total heat flux is the only heat flux that is measured directly in setups concerned with heat transfer. This total heat flux includes the radiative heat flux and the conductive heat flux through the spacers, the convective heat flux being prevented by operating in vacuum conditions. In the case of radiative heat transfer experiments, the following formula was used to deduce the radiative heat transfer from the total heat transfer measured by calorimetry:

$$Q_{rad} = Q_{total} - Q_{cond} \tag{1}$$

where $Q_{rad}$ is the radiative heat transfer and $Q_{total}$ is the total heat transfer measured in the setup. Therefore, the radiative heat flux values presented in the literature have been deduced using the total measured heat flux and the calculated conductive heat flux. In nearly every article concerned with plane-to-plane heat transfer, $Q_{cond}$ was calculated using Fourier's law or a more sophisticated FEM model. No experimental values for $Q_{cond}$ have been reported in experiments focusing on radiative heat transfer. Therefore, all published values of the radiative heat flux have been approximated. In the near-field regime, the impact of this approximation is probably not significant, considering the radiative heat transfer is greatly enhanced and theoretically far greater than conductive heat transfer. However, an experimental distinction between heat losses and radiative energy transfer has not yet been reported, highlighting a significant gap in current research.

In the case of thermionic devices, the electron flux is produced via the conversion of heat to electrons. Therefore, both radiative and conductive heat fluxes represent parasitic losses for TICs. This explains why the vacuum gap distance cannot be scaled down to nano scale sizes as the radiative heat transfer would be predominant through near field effects. To our knowledge, no experiments have been conducted on the relationship between near-field radiative heat flux and electricity production in TICs due to constraints in characterizing the fluxes independently.

## 4.2. THERMAL RESISTANCE OF INTERFACES

Typically, conduction through the spacers has been approximated by Fourier's law. These losses are directly dependent on the surface coverage (SC), thermal conductivity of the spacer material ($\kappa_{spacer}$) and length of the spacer (L):

$$Q_{cond} = h_{cond}.\Delta T = \frac{SC.\kappa_{spacer}.\Delta T}{L} \tag{2}$$

To calculate accurately the conduction through the spacers one would require complete knowledge of the height distributions and contact surface areas of the spacers, as well as the deformation of the sample when heat and external forces are applied [62]. Those parameters are not available experimentally and highly depend on the limitations imposed by the fabrication of the spacers. For example, several authors noted a strong disparity in the spacers' heights, which contributed to uncertainty in conduction losses [60], [64], [97]. The contact area used to determine the surface coverage has also been approximated often, most prevalently in cases of spherical spacers. The contact area of spherical spacers was almost always evaluated by using the cross-sectional area of a cylinder. Only Sabbaghi *et al.* [91] went beyond this approximation by implementing the Hertz model to estimate the contact area between the polystyrene spheres and emitter/receiver surfaces. Additionally, the value of the thermal conductivity of the spacer is often taken as tabulated in literature without considering the variation due to temperature changes as discussed previously. Therefore, Equation (2) should be considered an approximation of the conduction losses. Considering interfacial thermal contact resistances between surfaces in a TIC or NF-TPV, illustrated in the thermal circuit on the right side of Figure 8, represents a more precise approach to evaluating conduction losses through spacers. The total thermal resistance of the vacuum gap $R_{\mathrm{gap}}$ can be expressed as:

$$R_{\mathrm{gap}} = R_{\mathrm{e-s}}^{\mathrm{int}} + R_{\mathrm{spacer}}^{\mathrm{cond}} + R_{\mathrm{s-r}}^{\mathrm{int}} \tag{3}$$

where $R_{\mathrm{e-s}}^{\mathrm{int}}$ and $R_{\mathrm{s-r}}^{\mathrm{int}}$ are the interfacial thermal contact resistances between the emitter and the spacer and between the spacer and the receiver, respectively. $R_{\mathrm{spacer}}^{\mathrm{cond}}$ is the thermal resistance associated with conduction through the spacer; it is dependent on the thermal conductivity of the bulk material which is itself dependent on the temperature of the material:

$$R_{\mathrm{spacer}}^{\mathrm{cond}} = \frac{\mathrm{L}}{\kappa_{spacer}\mathrm{A}_{\mathrm{spacer}}} \tag{4}$$

In most articles, only $R_{\mathrm{spacer}}^{\mathrm{cond}}$ was considered when calculating the conduction losses. This approach is valid for long spacers where $R_{\mathrm{spacer}}^{\mathrm{cond}} \gg R_{\mathrm{e-s}}^{\mathrm{int}} + R_{s-r}^{int}$. However, when the spacer length decreases, $R_{\mathrm{gap}}$ cannot be directly approximated as $R_{\mathrm{spacer}}^{\mathrm{cond}}$. This approximation becomes even more inaccurate if the spacer material is characterized by a high thermal conductivity. Therefore, considering the thermal contact resistances at the device's interfaces would result in lower values of the conduction losses since $R_{gap}$ would increase per Equation (3).

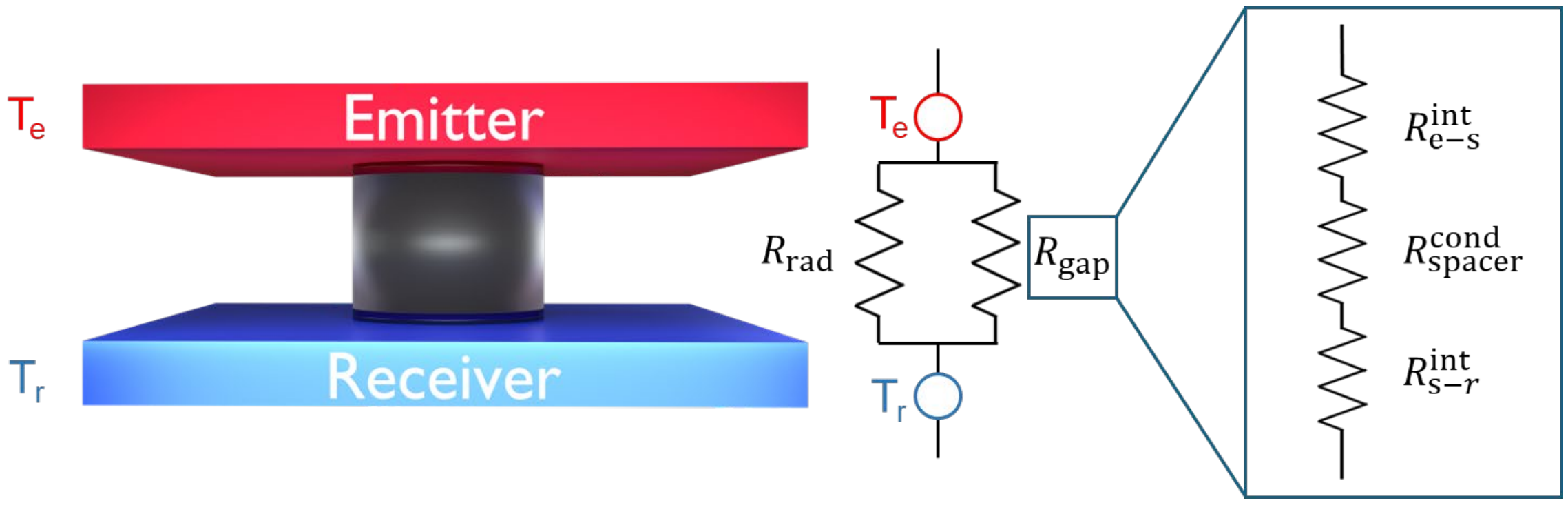

*Figure 8- Diagram (left) and corresponding thermal circuit (right) of a typical emitter/gap/receiver stack.*

If the contribution of the interfacial thermal resistances is taken into account, the effective conduction losses $Q_{\mathrm{cond,eff}}$ can be expressed as: $Q_{\mathrm{cond,eff}}=\Delta T/R_{\mathrm{gap}}$. This approach has been investigated to various degrees in several articles [22], [28], [30], [61], [84], [87], [97]. In two publications by Ito *et al.* [61], [87], the thermal contact resistance at the top surface of the truncated pyramid spacers, i.e., the thermal contact resistance between the emitter and the spacer, was obtained as a fitting parameter in their model. Ito *et al.* [61] modeled $R_{\mathrm{e-s}}^{\mathrm{int}}$, $R_{\mathrm{s-r}}^{\mathrm{int}}$ and $R_{\mathrm{spacer}}^{\mathrm{cond}}$ for a single spacer and found that the thermal contact resistances far outweighed the thermal resistance of the spacer structure alone, commenting that thermal contact resistances are vital in suppressing conduction losses through the spacers. Belbachir *et al.* [28] created an experimental setup for measuring the thermal resistance of the vacuum gap when various pressures ranging from 0.04 MPa to 0.37 MPa were applied. As expected, the thermal resistance of the entire vacuum gap decreased as the applied pressure increased due to the larger contact area formed at the interface between the spacers and the emitter/receiver. When the corresponding conduction losses were calculated using the measured $R_{\mathrm{gap}}$ value of 2.4 $K.W^{-1}$, as opposed to Equation (2), which only considers the thermal resistance through the spacer geometry, the magnitude of the conduction losses was cut in half [28], further indicating that Equation (2) overestimates conduction losses. Two years later, the thermal resistance measurements in [28] were repeated, but the width of the silicon dioxide spacers was varied. The thermal resistance of the entire gap increased as the spacer width decreased, which was expected due to the relationship between thermal resistance and area expressed in Equation (4) [84]. The authors emphasized that experimental measurements were essential to investigate thermal contact resistance, as the validity of a model can be affected by a change of temperature or applied pressure [28], [119]. An increase of contact pressure or force applied to the system can lower the interface's thermal resistance by increasing the contact area between the spacer and the surface. In some cases, neglecting thermal contact resistance was justified by the authors. De Sutter *et al.* [73] ignored because the SU-8 reflowed and filled any voids during the bonding process, creating good contact. Finally, Habibzadeh *et al.* [97] ignored $R_{\mathrm{s-r}}^{\mathrm{int}}$ because the SU-8 spacers were fabricated on the receiver, but $R_{\mathrm{e-s}}^{\mathrm{int}}$ was included in conduction loss calculations depending on the presence of graphene.

The most straightforward way of increasing interfacial thermal contact resistances has been implementing spherical spacers, since these spacers have a very low contact area [60], [81]. Removable spacer films, such as the corrugated films designed by Nicaise *et al.* [30] and Campbell *et al.* [40], are also promising and have been more reliable in size and ability to remain fixed during experiments than spheres. In both works, the thermal resistance of the corrugated spacer film was measured for various applied pressures ranging from 1 atm to 5.5 atm. Upon fitting the results to a model that considered surface roughness, Nicaise *et al.* [30] deduced that approximately one-third of the total thermal resistance could be attributed to the contact resistance between the spacer film and the emitter/receiver surfaces. This result further indicates that contact thermal resistances should not be neglected when evaluating conduction losses, especially in the case of unattached spacers since the lack of attachment results in higher contact resistances. In [30], the measured thermal resistance was also used to calculate an effective thermal conductivity of the alumina spacer structure. The result was ~5 $mW.m^{-1}.K^{-1}$, which was lower than SU-8 or other organic materials and on par with insulating aerogels. Later, in [40], the thermal resistance of a similar corrugated spacer film was further improved by utilizing alumina-hafnia alloys and alumina-hafnia-zirconia nanolaminate films as spacer materials as opposed to pure alumina, and by adding raised protrusions at the intersections of the hexagonal unit cells to increase the contact resistance. The use of multilayered ceramics proved useful in augmenting the overall thermal resistance compared to the pure alumina film because of interlayer interfaces. However, it should be noted that these multilayered films had a thermal resistance that was more sensitive to applied pressure than a pure alumina film.

### 4.3. CHALLENGES IN MEASURING THE GAP DISTANCE

When characterizing spacers, accurate determination of the vacuum gap distance is essential to extract the contribution of each flux to the total heat transfer that is measured. Several methods of characterizing gap distance have been employed in literature. Some authors relied on reflectometry and curve fitting to approximate the gap distance [61], [62], [64], [66], [86], [87], [88]. This method has proven to be precise, with an error of roughly 10 nm, but it is limited to nonmetallic films due to the opacity of metals. Therefore, reflectometry would be difficult to apply outside of plane-to-plane experiments [91]. Some publications have estimated the gap distance through capacitance measurements between two conductive surfaces [30], [40], [85], [90]. This approach is advantageous because measurements can be taken in-situ, but strong variations in gap distance have been observed depending on the magnitude of the applied force meant to correct bowing [85]. The simplest method has been approximating the gap distance as the height of the spacer; this has been employed by the majority of researchers [28], [48], [60], [63], [72], [73], [81], [82], [84], [89], [92], [94], [95], [96]. The height of the spacer has been measured via direct methods, such as electronic or confocal microscopes, and the gap distance was assumed to equal to the length of the spacer. However, this assumption is valid only if the height distribution of the spacers across the surface is homogeneous, and if the surface is almost perfectly planar. In the case of micro- or nanospheres used as spacers, this method proved to be a poor estimation of the gap size due to the large variation in the spheres' diameters [60], [64], [81], [91]. Furthermore, unlike the reflectometry and capacitance methods, approximating the gap distance as the spacer height completely ignores potential deformation of the spacers under applied forces used to correct any bowing. To rectify this problem, Sabbaghi *et al.* [91] fit the value obtained for the measured near-field radiative heat flux between to Si plates to a theoretical value obtained by fluctuational electrodynamics. Shi *et al.* [92], [93] relied on height measurements to determine the gap distance in conjunction with Hooke's law, which was implemented to calculate the displacement the SU-8 spacers would undergo due to the applied force on the stack. The gap distance obtained by this method was verified by fitting the measured values to computed values of the radiative heat transfer.

## 5. CONCLUSION

Thermionic and thermophotovoltaic energy converters represent an interesting alternative to current thermoelectric converters by enabling potentially higher efficiencies and output power densities. However, to achieve expected performance metrics, both technologies must feature controlled vacuum gaps. Spacers appear to be the most straightforward and scalable design solution for creating nano- and microscale gaps, but these structures introduce a pathway for parasitic conduction losses, unlike other methods such as positioner actuator stages. A wide variety of materials have been tested as candidates for spacers, with the most popular materials being silicon dioxide and organic photoresists. Despite their popularity in the literature, organic photoresists are limited to low-temperature applications and are consequently not feasible for actual devices. Nevertheless, gaps less than 100 nm have only been achieved using SU-8 spacers, indicating that research efforts have yet to achieve NF-TPV devices with very small gaps that are also operational at high temperatures.

For near-field thermophotovoltaics, most experiments achieved relatively low temperature differences between the emitter and receiver and there are only a few experimental devices including a TPV cell. Future publications should focus on exploring a wider range of temperature differences with spacers fabricated from materials more suitable to high-temperature applications, such as ceramics. For thermionics more devices with spacers have been fabricated, however, the impact of spacers on the electron flux and the overall efficiency of the system has not been explored. Spacer materials should be chosen according to their thermal and electrical conductivities, as well as their ability to withstand high temperatures and possible compression without breaking. In this regard ceramics and most prominently, multilayered ceramics, present an interesting pathway to reduce greatly the conduction losses between the emitter and the receiver.

The design of the spacers can be refined to further reduce the conduction losses, either by increasing the overall thermal resistance by boosting thermal contact resistance or by increasing the length of the spacers posts. For the first approach, using removable spacers has already been proven to be effective with corrugated films being a promising approach for thermionics and potentially for thermophotovoltaics if their size could be reduced. Increasing the spacers' length can be accomplished by embedding the posts in either the emitting or receiving surface to lengthen the conduction pathway. Although this method requires additional fabrication steps that are less straightforward than having spacers directly on one of the surfaces, it is maybe the simplest way to have a robust design with minimized conduction losses.

Combining several approaches to reducing conduction losses could result in even lower conduction losses than any one method alone. For example, using embedded spacers with maximized lengths fabricated from a material with very low thermal conductivity could result in minimal conduction losses. In this sense, the use of multilayered ceramics could be promising, as these materials have ultralow thermal conductivities and high resistance to elevated temperatures.

Characterization of the gap size and conduction losses should be considered in future research to fully understand the viability of spacers in micro- and nanoscale devices that must withstand multiple thermal cycles. Spacers have been applied to relatively small areas only, but they are scalable to larger areas, which is important in maximizing power output density. As such, understanding their impact on both types of devices should be at the forefront of research efforts, especially since there seems to be a tradeoff between minimizing conduction losses and maintaining mechanical robustness over a larger area. Exploring a wider range of spacer materials, specifically ceramics, to enable devices with higher emitter temperatures should also be prioritized.

## ACKNOWLEDGEMENTS

This work is part of the projects TED2021-131778B-C21 (funded by MCIN/AEI/10.13039/501100011033 and the European Union "NextGenerationEU"/PRTR) and PID2020-115719RB-C22 (funded by MICIU/AEI/10.13039/501100011033).

*Table 1- Key parameters collected to conduct the review. The surface coverage represents the total area of the emitter or receiver surface that is covered by spacers: $SC=N.A_{spacer})/A_{surface}$, where N is the number of spacers, $A_{spacer}$ is the contact area of one spacer with the surface onto which it is deposited or fabricated and $A_{surface}$ is the area of the surface on which the spacer is deposited/fabricated. Values denoted with an asterisk were stated explicitly in the corresponding research article, while values without an asterisk were calculated using the information present in the paper. A dash means that the information was not provided in the articles.*

| Article | Application | Type | Emitter | Receiver | Material of the spacer | Shape of the spacer | Distribution | Area hosting the spacer ($cm^{-2}$) | Surface coverage | Minimal gap distance (nm) | Maximal $T_{emitter}$ (K) | Maximal ΔT | Pressure (Pa) |
|---|---|---|---|---|---|---|---|---|---|---|---|---|---|
| Hu, Appl. Phys. Lett., 2008 [60] | TPV | heat transfer | $SiO_2$ | $SiO_2$ | Polystyrene | Spheres | Unattached | 1,27 | 4,96E-07 | 1000 | 380 | 83 | 8,50E-03 |
| Ito, Appl. Phys. Lett., 2015 [61] | TPV | heat transfer | $SiO_2$ | $SiO_2$ | $SiO_2$ | Truncated pyramids | On both sides | 1,63 | 8,00E-06 | 500 | 313 | 20 | 5,00E-03 |
| Watjen, Appl. Phys. Lett., 2016 [62] | TPV | heat transfer | n-doped Si | n-doped Si | $SiO_2$ | Cylindrical columns | — | 1,00 | 3,14E-06 | 200 | 337 | 35 | 3,00E-04 |
| Bernardi, Nat. Commun., 2016 [63] | TPV | heat transfer | undoped Si | undoped Si | $SiO_2$ | Cylindrical columns | Receiver | 0,25 | 3,14E-06 | 150 | 420 | 120 | 1,00E-04 |
| Ito, Nano letters, 2017 [87] | TPV | heat transfer | $SiO_2$ | $SiO_2$ with or without W-doped VO2 film | $SiO_2$ | Truncated pyramids | Both | 1,60 | 3,06E-05 | 370 | 370 | 45 | — |
| Lang, Sci. Rep., 2017 [64] | TPV | heat transfer | $SiO_2$ (fused or BK7) | $SiO_2$ (fused or BK7) | $SiO_2$ | Spheres | Unattached | 3,14 | 7,00E-06 | 150 | 300 | 7 | 1,00E-03 |
| Yang, Nat. Commun., 2018 [88] | TPV | heat transfer | Graphene on undoped or n-doped Si | Graphene on undoped or n-doped Si | AZ Photoresist | Cylindrical columns | — | 4,00 | 1,96E-05 | 430 | 353 | 50 | 6,67E-04 |
| DeSutter, Nat. Nanotechnol., 2019 [73] | TPV | heat transfer | p-doped Si | p-doped Si | SU-8 | Cylindrical columns | Emitter | 0,25 | 1,00E-04* | 110 | 400 | 100 | 5,00E-04 |
| Tang, ACS Photonics, 2020 [89] | TPV | heat transfer | SiC or p-doped Si | p-doped Si | $SiO_2$ | Cylindrical columns | Receiver | 0,25 | 1,40E-06* | 150 | 380 | 80 | 1,00E-04 |
| Ying, ACS Photonics, 2020 [90] | TPV | heat transfer | p-doped Si | p-doped Si | SU-8 | Cylindrical columns | Receiver | 0,70 | 5,25E-06 | 190 | 385 | 85 | 1,00E-01 |
| Sabbaghi, J. Appl. Phys., 2020 [91] | TPV | heat transfer | Al-coated doped Si | Al-coated doped Si | Polystyrene | Spheres | Unattached | 0,25 | 4,10E-06 | 215 | 361 | 65 | 1,00E-01 |
| Shi, Adv. Mater., 2021 [92] | TPV | heat transfer | Graphene/SU-8 heterostructures on $SiO_2$/Si | Graphene/SU-8 heterostructures on $SiO_2$/Si | SU-8 | Cylindrical columns | Receiver | 0,09 | 2,79E-04 | 55 | 328 | 25 | 2,80E-05 |
| Lu, Small, 2022 [94] | TPV | heat transfer | hBN, graphene, graphene/hBN or graphene/hBN/graphene multilayers on undoped Si | hBN, graphene, graphene/hBN or graphene/hBN/graphene multilayers on undoped Si | AZ Photoresist | Cylindrical columns | Receiver | 1,00 | 2,04E-05 | 400 | 323 | 40 | 1,00E-03 |
| Shi, Nano Lett., 2022 [93] | TPV | heat transfer | Graphene/SU-8 heterostructures on $SiO_2$/Si | Graphene/SU-8 heterostructures on $SiO_2$/Si | SU-8 | Cylindrical columns | Receiver | 0,09 | 2,79E-04 | 81 | 308 | 5 | 2,80E-05 |
| Li, Appl. Phys. Lett., 2024 [95] | TPV | heat transfer | W-coated Si, ITO/W-coated Si, and $SiO_2$/W multilayers on Si | GaSb | $SiO_2$ | Cylindrical columns | Emitter | 1,00 | 2,51E-06 | 200 | 383 | 100 | 1,00E-03 |
| Li, J Quant Spectrosc Radiat Transfer, 2024 [96] | TPV | heat transfer | $SiO_2$ on undoped Si | $SiO_2$ on undoped Si | $SiO_2$ | Cylindrical columns | Emitter | 0,25 | 4,52E-06 | 302 | 337 | 42 | 8,00E-04 |
| Habibzadeh , ACS Photonics, 2024 [97] | TPV | heat transfer | SiC or LiF or LiF with Graphene layer | SiC | SU-8 | Cylindrical columns | Receiver | 2,25 | 1,44E-05 | 120 | 357 | 60 | 1,20E-03 |
| DiMatteo, Appl. Phys. Lett., 2001 [48] | TPV | TPV device | Si | InAs PV cell | $SiO_2$ | Cylindrical columns | Emitter | 0,05 | — | 1000 | 408 | — | 5,33 |
| DiMatteo, AIP Conf. Proc. 738, 2004 [85] | TPV | TPV device | Si | InGaAs PV cell | $SiO_2$ | Hollow cylinders | Emitter | 0,20 | — | 120 | 1123 | — | — |
| Inoue, Nano Letters, 2019 [66] | TPV | TPV device | Undoped Si | InGaAs PV cell | Si | Cylindrical columns | Receiver | 0,0025 | — | 140 | 1040 | > 700 | 1,00E-03 |
| Inoue, ACS Photonics, 2021 [86] | TPV | TPV device | Undoped Si | InGaAs PV cell | Si | — | Receiver | 0,01 | — | 140 | 1200 | 900 | 1,00E-03 |
| Selvidge, Adv. Mat., 2024 [72] | TPV | TPV device | GaAs | InAs PV cell | GaAs | Rectangular posts | Emiiter | 0,28 | 8,93E-06 | 150 | 733 | 470 | 1,00E-03 |
| Yao, Journal of heat transfer, 2009 [74] | TIC | spacer characteriz ation | Si | Au-coated Si | SU-8 | Semi-circular hollow columns | Emitter | 1,00 | 5,97E-04 | 10000 | 353 | 30 | 1 |
| Belbachir, Microsyst. Tech., 2016 [84] | TIC | spacer characteriz ation | SiC | Pt -coated Si | $SiO_2$ | Truncated pyramids | Receiver | 0,84 | 1,71E-02 | 10000 | 373 | 80 | 1,01E+05 |
| Nicaise, Microsystems & Nanoeng., 2019 [30] | TIC | spacer characteriz ation | Si | Si | $Al_2O_3$ | U-beam ribs arranged in hexagons | Unattached | 0,90 | 2,17E-03 | 3000 | 473 | — | 4,00E-04 |
| Campbell, J. Microelectro. Sys., 2020 [40] | TIC | spacer characteriz ation | W-coated Si | W-coated Si | $Al_2O_3$ | U-beam ribs arranged in hexagons | Unattached | 1,25 | 7,23E-04 | 1800 | 403 | 40 | 1,33E-02 |
| | TIC | TIC device | Mo | Mo | $Al_2O_3$ | U-beam ribs arranged in hexagons | Unattached | 3,14 | — | 2300 | 1280 | 460 | — |
| Beggs, Adv. Energy Convers., 1963 [35] | TIC | TIC device | SrCaO/Pt-coated W | BaSrO-coated W | Mo | Sheets | Unattached | — | — | 5588 | 1423 | 500 | — |
| Fitzpatrick, IECEC-93, 1993 [76] | TIC | TIC device | W-Ta-Re alloy monocrystal | Mo-Nb alloy monocrystal | $Al_2O_3$ | — | Receiver | — | — | 9500 | 1306 | 509 | 20 |
| King, AIP Conference proc., 2001 [78] | TIC | TIC device | BaO/SrO/CaO/W-coated sapphire | BaO/SrO/CaO/W-coated $SiO_2$ | $SiO_2$ | unspecified | Receiver | 0,84 | — | 15000 | 1170 | 450 | vacuum condition but no details provided |
| Lee, Microsystems Workshop, 2012 [79] | TIC | TIC device | poly-SiC | Si | $SiO_2$ | Rectangular posts | Receiver | 0,0025 | — | 10000 | 1650 | — | 1,33E-04 |
| Littau, Phys. Chem. Chem. Phys., 2013 [81] | TIC | TIC device | Ba-activated W | W-coated Si | $Al_2O_3$ | Spheres | Unattached | 0,32 | — | 11000 | 1473 | — | 6,67E-04 |
| Lee, J. Microelect. | TIC | TIC device | BaO/SrO/CaO/W- | Si | $SiO_2$ | Rectangular posts | Receiver | 0,0025 | — | 5000 | 1400 | > 1000 | 1,33E-04 |
| Sys., 2014 [80] | | | coated poly-SiC | | | | | | | | | | |
| Belbachir, J. Micromech. Microeng., 2014 [28] | TIC | TIC device | SiC | Pt -coated Si | $SiO_2$ | Truncated pyramids | Receiver | 0,84 | 1,00E-01 | 10000 | 1103 | 462 | 1,00E-03 |
| Bellucci, Energy Tech., 2021 [82] | TIC | TIC device | W | GaAs | $ZrO_2$ | Cylindrical columns | Receiver | 0,41 | 3,20E-04* | 3000 | 1673 | 1120 | 5,00E-06 |
| Trucchi, Adv. Energy Matls, 2018 [83] | Hybrid TIC-TEC | TIC device | n-doped diamond film on HfC | Mo | $Al_2O_3$ | Ring | Unattached, but embedded in receiver | | — | 100000 | 1029 | 521 | 1,00E-04 |